\documentclass[twocolumn,showpacs,preprintnumbers,amsmath,amssymb]{revtex4}
\usepackage{amsfonts}
\usepackage{amsmath}
\usepackage{graphicx}
\usepackage{dcolumn}
\usepackage{bm}
\usepackage{overpic}
\usepackage{booktabs}

\begin{document}
\title{Dynamics on Spatial Networks and the Effect of Distance Coarse graining}
\author{An Zeng, Dong Zhou, Yanqing Hu\footnote{yanqing.hu.sc@gmail.com}, Ying Fan, Zengru Di}

 \affiliation{Department of Systems Science, School of Management and Center for Complexity
 Research, Beijing Normal University, Beijing 100875, China}

\date{\today}

\begin{abstract}
Very recently, a kind of spatial network constructed with power-law
distance distribution and total energy constriction is proposed.
Moreover, it has been pointed out that such spatial networks have
the optimal exponents $\delta$ in the power-law distance
distribution for the average shortest path, traffic dynamics and
navigation. Because the distance is estimated approximately in real
world, we present an distance coarse graining procedure to generate
the binary spatial networks in this paper. We find that the distance
coarse graining procedure will result in the shifting of the optimal
exponents $\delta$. Interestingly, when the network is large enough,
the effect of distance coarse graining can be ignored eventually.
Additionally, we also study some main dynamic processes including
traffic dynamics, navigation, synchronization and percolation on
this spatial networks with coarse grained distance. The results lead
us to the enhancement of spatial networks' specifical functions.


\end{abstract}



\maketitle

\section{Introduction}
The research on complex networks has been one of the most active
fields not only in physics but also in other various disciplines of
natural and social sciences\cite{RMP47,Nature440,Science509,RMP75}.
In traditional statistical mechanics, interaction mainly exists
between neighboring elements. By introducing the complex topology of
the networks, the whole system can emerge some new properties such
as small-world\cite{Nature440}, scale-free degree
distribution\cite{Science509} and community structures\cite{RMP75}.
However, the spatial property is of great significance as well,
which makes the interaction between nodes go beyond the neighboring
effect but under the restraint of their underlying geographical
site. This property matters much in lots of empirical networks
including neural network\cite{Complexity56}, communication
networks\cite{PRE015103}, the electric-power grid\cite{PRE025103},
transportation systems\cite{PNAS7794,PRE036125,PhysicaA109} and even
social
networks\cite{Social187,PhysicaA5317,PNAS11623,arxiv1802,arxiv1332}.
Generally, the geography information of the nodes and the distance
between nodes in these networks would determine the characteristics
of the network and play an important role in the dynamics happening
in the network.

About the networks embedded in the geographical space, many works
has been
done\cite{FC1994,PNAS13382,PRE037102,PRE026118,PhysicaA853,EPJB1434,PRE016117}.
The first category is focusing on the spatial distribution of the
nodes of these empirical spatial networks\cite{FC1994,PNAS13382}.
Specifically, networks with strong geographical constraints, such as
power grids or transport networks, are found with fractal
scaling\cite{FC1994}. Besides, others researchers discussed the
small-world behavior and the scale-free networks in Euclidean
space\cite{PRE037102,PRE026118}. For example, when supplementing
long range links whose lengths are distributed according to
$q(l)\propto l^{-\alpha}$ to D-dimensional lattices, Sen, Banerjee
and Biswas conjectured that the two transition points from random
networks and Regular networks to the networks with small-world
effect in any dimension are: $\alpha=D$ and $\alpha=D+1$
respectively\cite{PRE037102}. Also, Xulvi-Brunet and Sokolov
constructed an growing network model by $\prod\limits_{i\rightarrow
j}\propto k_{i}l_{ij}^{-\alpha}$, where $l_{ij}$ is the distance
between $i$ and $j$. Numerical simulations have shown that for
$\alpha<1$ the degree distribution is a power-law distribution and
for $\alpha>1$ it is fitted by a stretched
exponential\cite{PRE026118}. In addition, some researchers also
introduced some ways to model the empirical geographical
networks\cite{PhysicaA853,EPJB1434,PRE016117,arxiv1332}.

However, few of these former works in spatial networks are related
to the total cost restraint. In fact, The total cost is very
important when designing these real spatial networks. Because the
longer a link is, the more it will cost. Very recently, some
researches took this aspect into
account\cite{EPL58002,PRL018701,arxiv1802}. In \cite{EPL58002},
based on a regular network and subject to a limited cost $C$, long
range connections are added with power-law distance distribution
under the probability density function (PDF) $P(r)=ar^{-\delta}$.
Some basic topological properties of the network with different
$\delta$ are studied. It is found that the network has the minimum
average shortest path when $\delta=2$ in one-dimensional spatial
networks, In addition, the authors investigated a classic traffic
model on this model networks. It is found that $\delta=1.5$ is the
optimization value for the traffic process on the spatial networks.
In \cite{PRL018701}, pairs of sites $ij$ in 2-dimensional lattices
are randomly chosen to receive long-range connections with
probability $Pr(u,v)$ proportional to $r_{uv}^{-\alpha}$. With the
total energy restriction, G. Li et cl. found $\alpha=3$ is
corresponding to the minimum average shortest path. Moreover, they
claimed that the optimal value for navigation is $\alpha=3$ in
2-dimensional spatial networks and $\alpha=2$ in 1-dimensional
ones\cite{explain}.

In many empirical researches, especially the study on traffic
networks, the distances are estimate
approximately\cite{PNAS7794,PRE036125,PhysicaA109,PhysicaA5639}.
That is to say scientists incline to regard a range of distance as a
typical value. This procedure named distance coarse graining should
also be studied in the spatial network models. Actually, the coarse
graining process has been discussed since long times
ago\cite{PRL168701,PRL038701,PRL174104,PhysicaA5639}. All the
related works focus on how to reduce the size of the networks while
keep other properties unchanged such as degree distribution, cluster
coefficient, degree correlation\cite{PRL168701}, random
walks\cite{PRL038701} and synchronizability\cite{PRL174104}. On the
other hand, the probability distribution of these long
range-connections is chosen as $P(r)=ar^{-\delta}$ in the spatial
network. It means most of the connections are short while a few
connections are relatively long. However, each node has only two
neighbors. When the total cost constraint is chosen as a certain
large value in the binary network, the network can not provide
enough short long-range connections. In this paper, we use the
distance coarse graining to solve this problem. Specially, we will
study how the distance coarse graining affects the topology and
dynamical process in the spatial network model. The result shows
that the $\delta$ for minimum average shortest path, optimal traffic
process\cite{PRE026125,PRE046108} and
navigation\cite{Nature845,PRE017101,PRL238703,PhysicaA109} will
shift to smaller value. And the more we coarse grain the distance,
the more significantly the optimal $\delta$ will shift.
Interestingly, when the network is large enough, the effect of
distance coarse graining can be ignored. In other aspect, how the
dynamic processes perform in spatial networks is an interesting
problem but hasn't received enough attention. Investigating the
dynamics on the spatial networks can not only lead to enhancement of
the function of the spatial network but also provide us with a
better understanding of it. Here, we study two more main dynamic
processes and find synchronizability\cite{EPL48002,PR93} can also be
optimized by a typical $\delta$ while there is no optimal $\delta$
for percolation\cite{RMP1275} in such spatial network model.

\section{Generating Binary Spatial Networks with Coarse Grained Distance}
The model network is embedded in a $k$-dimensional regular network.
The long range connections is generated from a power-law distance
distribution. A total cost $C$ is introduced to this network model.
Every edge has a cost $c$ which is linear proportion to its distance
$r$. For simplification, the edge connecting node $i$ and $j$ cause
a cost represented by its length $r_{ij}$ in the model.

According to many empirical studies, the distance obeys the
power-law
distribution\cite{IPSJ155,GeoJournal102,Social187,PhysicaA5317,PNAS11623}.
Specially, for Japanese airline networks, even there is an
exponential decay in domestic flights, the distance distribution
follows power-law when international flights are
added\cite{IPSJ155}. For the U.S. intercity passenger air
transportation network, the distribution of the edge distance has a
power-law tail with the exponent
$\delta=2.20\pm0.19$\cite{GeoJournal102}. Additionally, authors in
ref.\cite{Social187,PhysicaA5317,PNAS11623} found $\delta=1$ for
social systems like the mobile phone communication networks.
Therefore, the probability distribution of these long range
connections here is chosen as $P(r)=ar^{-\delta}$ (PDF) in the
spatial model. In order to coarse grain the distance, when adding a
long range link to the network, we divide the distance into many
continuous parts in which all the distances are considered as one
typical value. So the spatial network is constructed as following:

\begin{enumerate}
\item $N$ nodes are arranged in a $1$-dimensional lattice. Every node is connected with its nearest neighbors which
can keep every node reachable. Additionally, between any pair of
nodes there is a well defined lattice distance.
\item Set the approximate interval $W$ of coarse graining process. For
each node, divide the distance between it and other nodes into
$\frac{N_{max}}{W}$ parts ($N_{max}$ is the largest distance between
any nodes in the initial network), the $m$th part are
$(m-1)W+1$,$(m-1)W+2$,...,$mW$.
\item A node $i$ is chosen randomly, and a certain distance
$r (2\leq r\leq N_{max}$) is generated with probability
$P(r)=ar^{-\delta}$, where $a$ is determined from the normalization
condition $\sum^{N_{max}}_{r=2}P(r)=1$.
\item Find the part that the distance $r$ belongs to, say $m$th part here. Then, one of the nodes in this part is
picked randomly, for example node $j$. An edge between nodes $i$ and
$j$ is created if there exists no edge between them yet.
\item  After step $4$, a certain cost $r_{ij}$ is generated. Repeat step $3$ and $4$ until the total
cost reaches $C$.
\end{enumerate}

Obviously, there are two significant features of the spatial
network: the power-law distribution of the long range connections in
the network and the restriction on total energy. Firstly, we should
determine how to choose an appropriate $W$ under different total
cost $C$. In the spatial network, the probability distribution of
these long range-connections is chosen as $P(r)=ar^{-\delta}$. It
means most of the connections are short while a few connections are
relatively long. However, each node has only two neighbors. When the
total cost reaches a certain value in the binary network, the short
range part of the network would become nearly full connected and can
not provide further short long-range connections. Consequently,
binary spatial network can not be generated directly without the
distance coarse grained. Clearly, if $W$ is too small, the spatial
network model still can not provide enough short long-range
connections. On the contrary, because all the nodes in one distance
coarse graining part are regarded as the same, if $W$ is too large,
too many nodes are considered in one distance coarse graining part
so that the power-law distance distribution will be destroyed
significantly.

In order to deduce the formula of appropriate $W$, we consider an
extreme condition with $\delta$ equaling to a very large positive
value. Under this circumstance, all the long-range connections are
short. So all of them will locate in the first distance coarse
graining part of each node. Consequently, the total energy is
$C=(2+3+4+...+W)N$. For $C=cn$, $2+3+4+...+W=c$ where $c$ is the
average energy on each node. After simplification, $W^{2}+W-2-2c=0$.
So we can get $W=\frac{\sqrt{9+2c}-1}{2}$. Generally, $W$ only has
to obey $W\geq [\frac{\sqrt{9+2c}-1}{2}]_{floor}$ where the
$[\cdot]_{floor}$ represents the operation of rounding downward. In
this paper, we choose $W=4$ when $c=10$, $W=7$ when $c=30$ and $W=9$
when $c=50$.

As mentioned above, the binary spatial network can not be generated
directly. In ref\cite{EPL58002}, to get the binary spatial network,
authors project the weighted spatial network into unweighted one by
imposing all the weight of the existing links to $1$, this will lead
to losing total energy. On the contrary, when the distance is coarse
grained, the network can provide enough short-term links. Under this
circumstance, the binary spatial network can be generated
independent from the weighted one. In Fig.1, we compare the total
energy of these two different binary spatial networks. Clearly, only
the network with coarse grained distance can satisfy the total
energy limit as the standard value.

\begin{figure}
  \center
  \includegraphics[width=6cm]{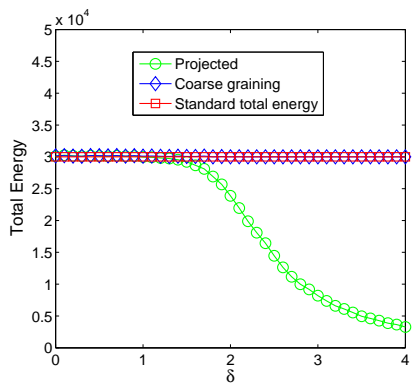}
  \caption{(Color online) The total energy of two different binary spatial networks. The cycle represents the projected binary spatial networks in ref.[11]. The lozenge stands for the binary spatial network with coarse grained distance with $W=7$.
  The square is the standard total energy which equals $C=n\times c=1000\times 30=30000$. The results are under 100 independent realization.}
\end{figure}

For the power-law distance distribution, just like we discussed
above, there is no doubt that it will be destroyed when the distance
is coarse grained. The results for distance distribution is reported
in Fig.2. We can see that the distance distribution in projected
binary spatial networks are almost the same as the standard distance
power-law distribution. On the contrary, in networks with coarse
grained distance, the distribution is different from the standard
one. This is reasonable, because the node in each coarse grained
part are regarded homogeneous. Consequently, the distribution plot
is also divided into some continuous parts in which the distance
distribution is normal. However, when we estimate the distance with
higher scale, which means marking all the distance from $(m-1)W+1$
to $mW$ as a specific value $(m-1)W+1$, the distance distribution
become exactly the same with the standard power-law, see the green
line in Fig.2.

\begin{figure}
  \center
  \includegraphics[width=6cm]{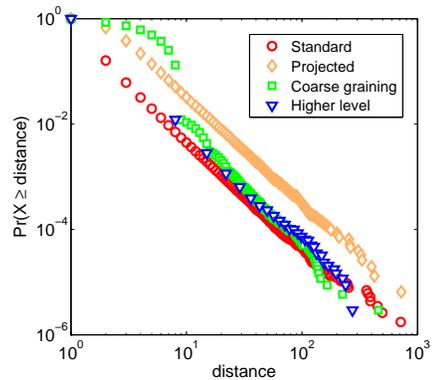}
  \caption{(Color online) Distance distribution of two different spatial networks. The lozenge represents result of the projected binary spatial networks in ref.[11].
  The square stands for the distance distribution under original scale of the binary spatial network with coarse grained distance $W=7$.
  The triangle is the distance distribution under higher scale (See the text) of the coarse grained network.
  The cycle is the standard distance power-law distribution with given $n=10000$, $c=30$ and $\delta=3$. The results are under 10 independent realization.}
\end{figure}

One of the most important results in the former works about spatial
networks is that the minimum shortest path happens at $\delta=2$
regardless of the total energy\cite{EPL58002,PRL018701}. Here, we
investigate how the optimal $\delta$ changes as the distance coarse
graining procedure. Fig.3 shows the result for three different level
of distance graining. Obviously, the larger $W$ we choose to coarse
grain the distance, the more severely the optimal point shifts to a
lower value. However, as the size of the network is getting bigger,
the shifting effect of the distance coarse graining becomes less
significant. That is to say, notwithstanding the coarse graining in
distance, the $\delta$ for the minimum shortest path still equals to
$2$ in large networks.

\begin{figure}
        \includegraphics[width=4.2cm]{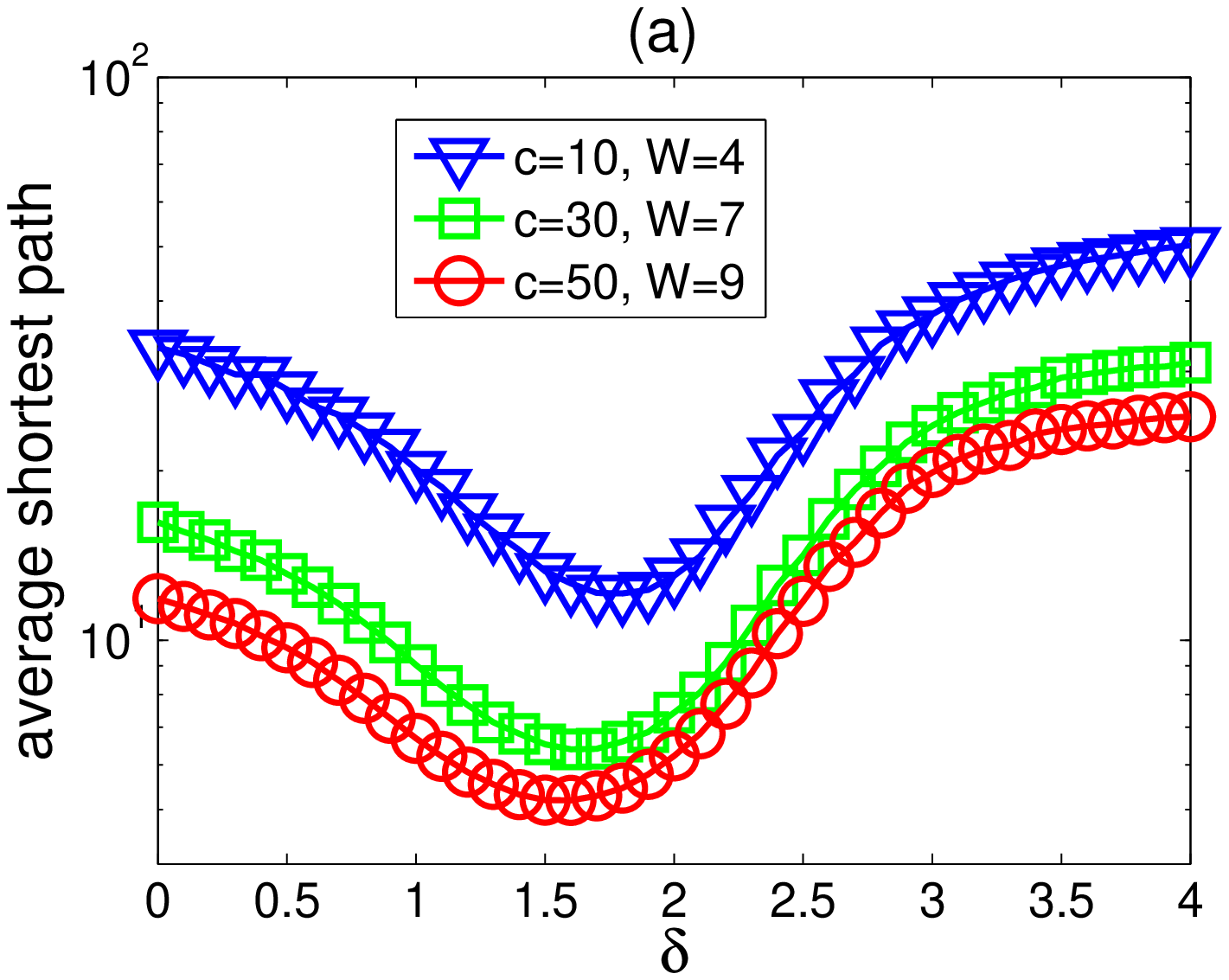}
        \includegraphics[width=4.2cm]{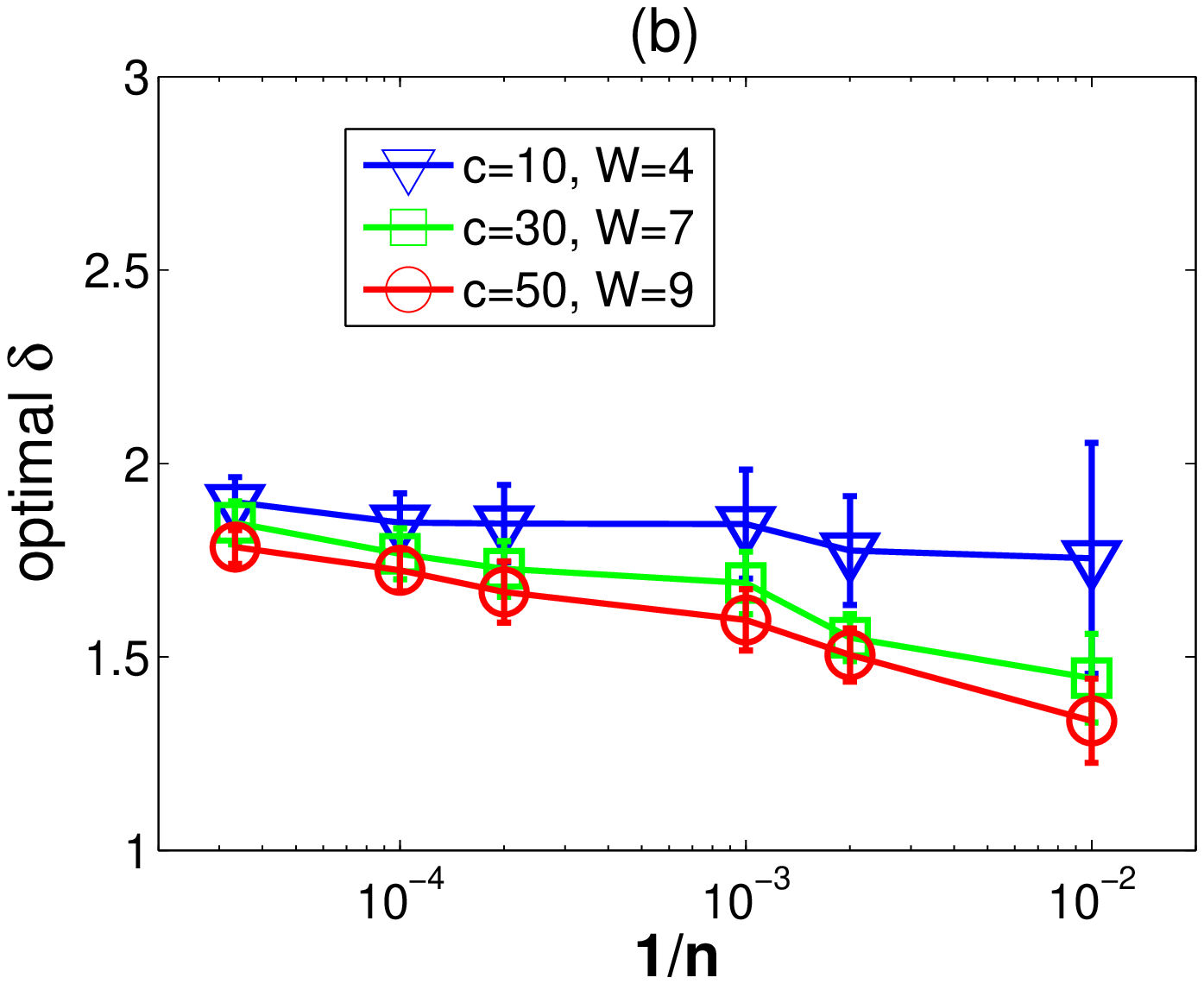}
\caption{(Color online) The result for how the optimal $\delta$
changes as the distance coarse graining procedure. (a) is the
average shortest path length for different $\delta$ under different
coarse graining scale, here $N=1000$. (b) is how the $\delta$ for
minimum average shortest path changes when the network getting
larger. The results are under 100 independent realization.}
\end{figure}

\section{The dynamics on spatial networks and the effect of distance coarse graining}
Dynamics on spatial networks is an interesting topic, for example,
it can help us obtain the principle to design the optimal
transportation networks\cite{PRL018701}. In this section, some main
dynamics process will be studied on the spatial networks.
Furthermore, most of the time, people tend to coarse grain the
distance when designing the real networks, especially these
transport networks. So understanding how the distance coarse
graining affects the function of the network can be not only of
great interest but also useful. So we will also discuss the effect
of distance coarse graining on the function of spatial networks.
Firstly, based on the results in the former
works\cite{EPL58002,PRL018701}, we know that the spatial properties
of the network will result in optimal $\delta$ for navigation and
traffic process respectively. So how the optimal $\delta$ shifts
with the distance coarse graining procedure will be investigated.
Secondly, we will study the synchronizability in this so-called
binary spatial network with distance coarse grained. Finally, the
percolation performance in this binary spatial network will be
studied as well.

\subsection{The Effect on Traffic Process}
In traffic process\cite{PRE026125}, All the nodes embedded on the
spatial network are treated as both hosts and routers. Every node
can deliver at most $D$ packets one step toward their destinations.
At each time step, there are $R$ packets generated homogeneously on
the nodes in the system. The packets are delivered from their own
origin nodes to destination nodes by special routing strategy. There
is a critical value $R_{c}$ which can best reflect the maximum
capability of a system handling its traffic. In particular, for
$R<R_{c}$, the numbers of created and delivered packets are
balanced, leading to a steady free traffic flow. For $R>R_{c}$,
traffic congestion occurs as the number of accumulated packets
increases with time, simply because the capacities of the nodes for
delivering packets are limited.

In fact, the whole traffic dynamics can be represented by analyzing
the largest betweenness of the network\cite{PRE046108}. The
betweenness of a node is the number of shortest path passing through
this node. Note that with the increasing of parameter $R$ (number of
packets generated in every step), the system undergoes a continuous
phase transition to a congested phase. Below the critical value
$R_{c}$, there is no accumulation at any node in the network and the
number of packets that arrive at node $i$ is $Rg_{i}/N(N-1)$ on
average. Therefore, a particular node will collapse when
$Rg_{i}/N(N-1)>D_{i}$, where $g_{i}$ is the betweenness coefficient
and $D_{i}$ is the transferring capacity of node $i$. Therefore,
congestion occurs at the node with the largest betweenness. Thus
$R_{c}$ can be estimated as $R_{c}=D_{i}N(N-1)/g_{max}$, where
$g_{max}$ is the largest betweenness coefficient of the network.

Here, we study the traffic dynamics in the spatial network with
coarse grained distance. The results are given in Fig.4(a) and (b).
In Fig.4(a), it is quite obvious that there exists an optimal
$\delta$ for $Rc$, which means the transport capacity reaches its
maximum in this spatial networks. From Fig.4.(b), we can clearly see
that this optimal $\delta$ gets closer to $1.5$ gradually as the
size of the network becomes larger. Actually, even losing the total
energy,  the projected spatial network also has the optimal $\delta$
for traffic process equalling to $1.5$ in ref.\cite{EPL58002}. This
implies that the spatial property play an dominative part in the
traffic dynamics.

\begin{figure}
        \includegraphics[width=4.2cm]{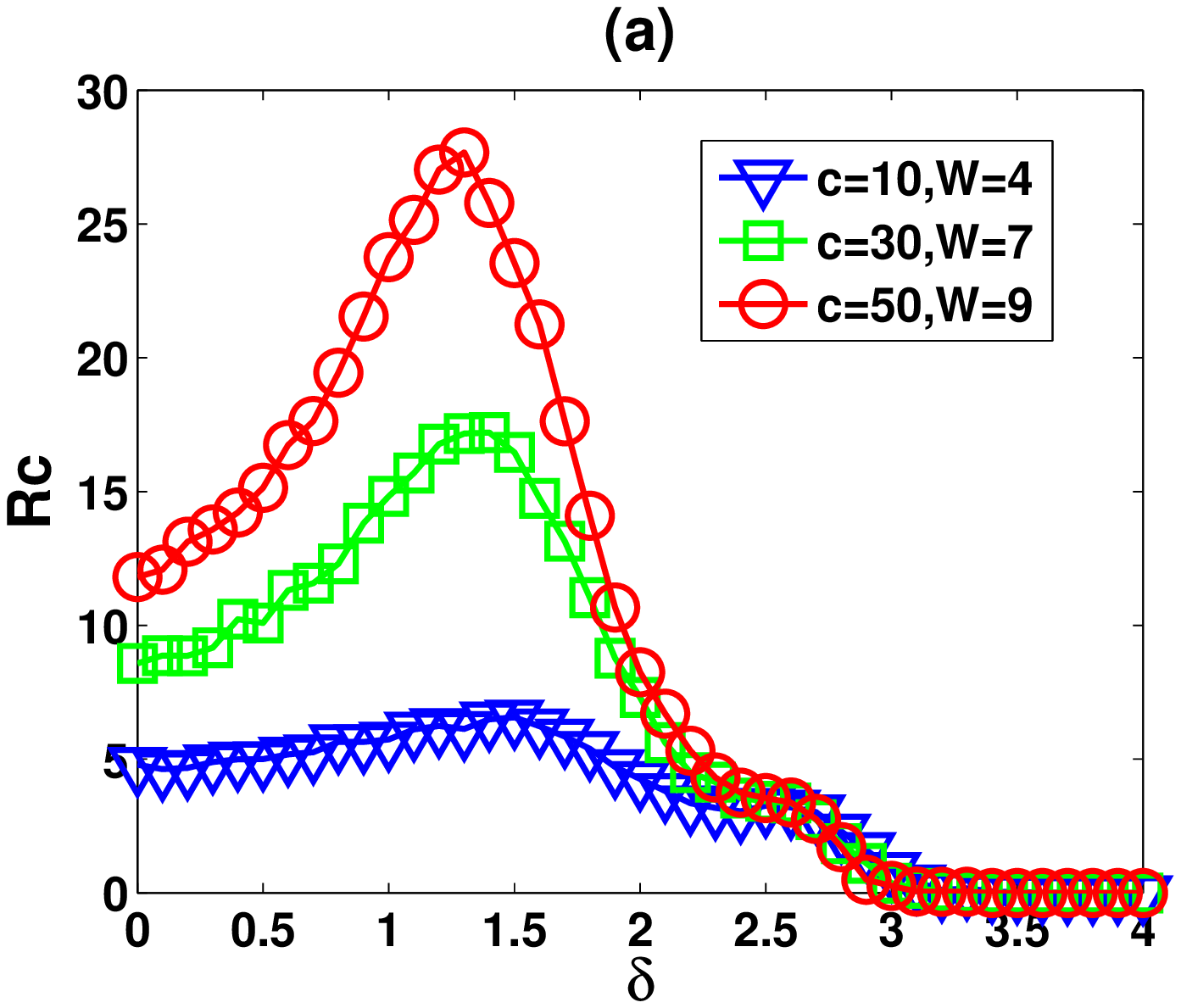}
        \includegraphics[width=4.2cm]{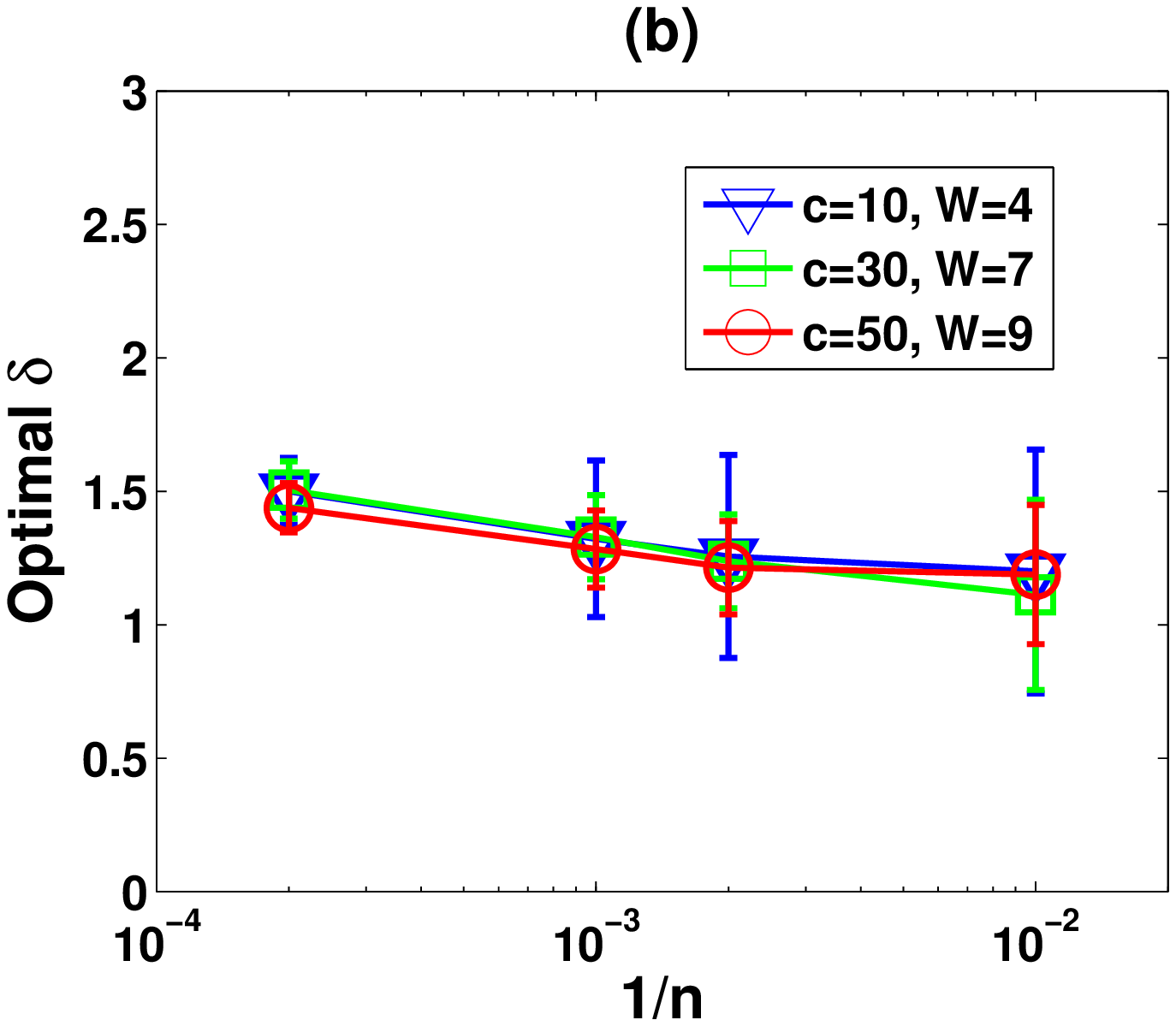}
        \includegraphics[width=4.2cm]{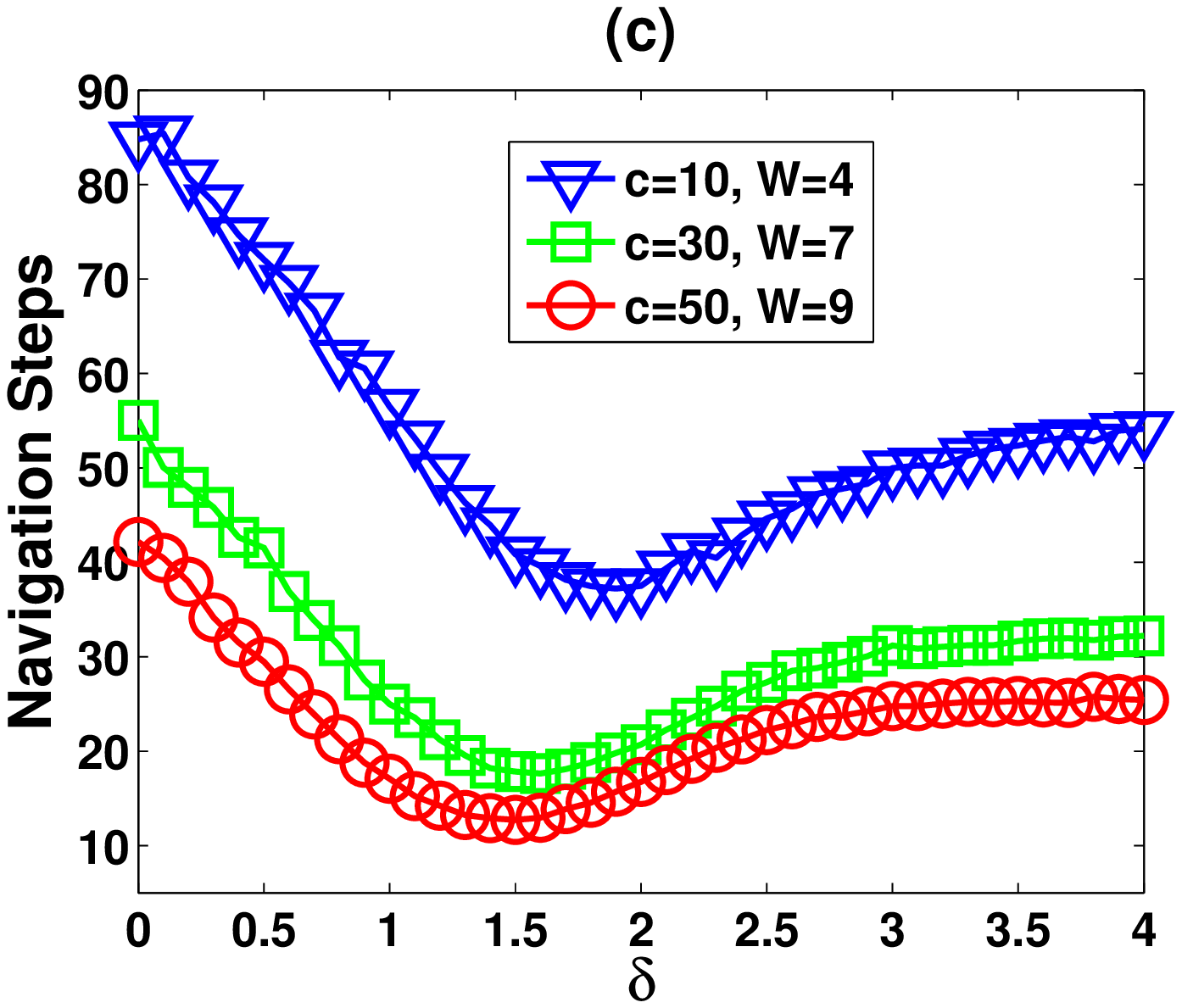}
        \includegraphics[width=4.2cm]{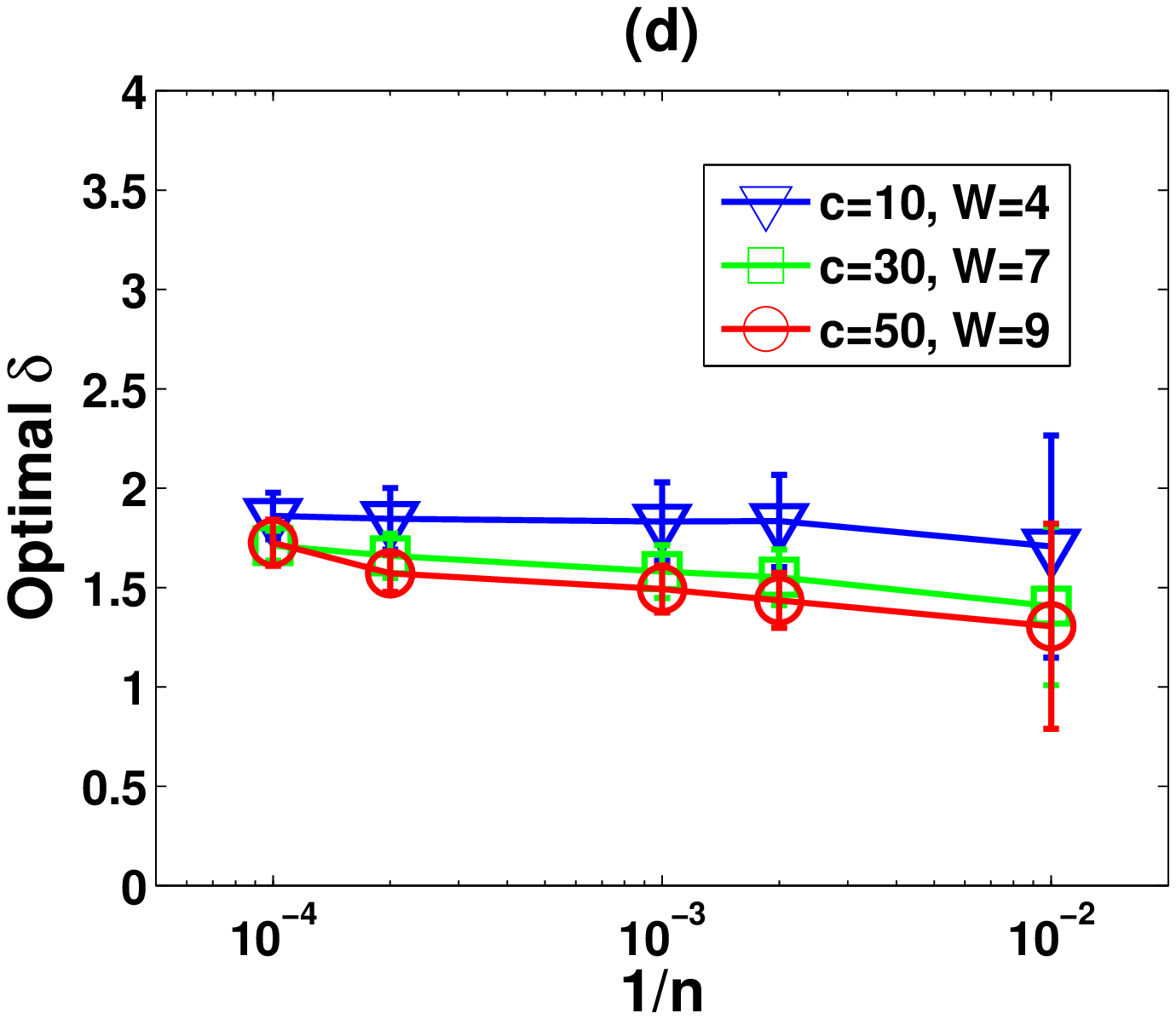}
\caption{(Color online) The effect of distance coarse graining on
the function of the binary spatial networks. (a) is the critical
value $Rc$ in traffic process for different $\delta$ under different
coarse graining scale, where $N=1000$. (b) shows how the optimal
$\delta$ for traffic process changes when the network gets larger.
(c) is the navigation steps for different $\delta$ under different
coarse graining scale, where $N=1000$. (d) shows how the optimal
$\delta$ for navigation changes when the network gets larger. The
results are under 50 independent realization.}
\end{figure}

\subsection{The Effect on Navigation}
Another aspect we are going to investigate here is navigation. In
fact, the navigation process in networks is based on local
information, which is different from the shortest path with global
information. Hence, the navigation reflects another ability of the
networks. According to ref.\cite{PRL018701}, we choose the
navigation strategy as the greedy algorithm\cite{Nature845} in this
paper. In the former works, Kleinberg found that $\alpha=2$ for
$Pr(u,v)\propto r_{uv}^{-\alpha}$ is the optimal value in the
navigation with the greedy algorithm in 2-dimensional spatial
networks without total energy limit\cite{Nature845}. When adding the
energy restriction, G. Li et cl. found that the optimal value is
$\alpha=3$ in 2-dimensional spatial networks and $\alpha=2$ in
1-dimensional ones\cite{PRL018701}.

Here, though we choose the PDF of the distance distribution as
$P(r)=ar^{-\delta}$, $\delta$ equals to $\alpha$ in one-dimensional
space\cite{explain}. What interest us most is that how this optimal
value performs when the distance is coarse grained. In Fig.4 (c) and
(d), the results are given. The optimal $\delta$ shifts to a smaller
value as the coarse grained interval $W$ gets larger. However, when
the size of the networks is large enough, the optimal $\delta$ will
come back to $2$ even if the distance gets coarse grained, as show
in Fig.4(d).

\subsection{The Effect on Synchronizability}

Furthermore, we will study the synchronizability in this so-called
binary spatial network with distance coarse grained. The
synchronization is a universal phenomenon emerged by a population of
dynamically interacting units. It plays an important role from
physics to biology and has attracted much attention for hundreds of
years.

In the former works, the analysis of Master Stability Function (MSF)
allows us to use the eigenratio $R=\lambda_{N}/\lambda_{2}$ of the
Laplacian matrix to represent the synchronizability of a
network\cite{PR93}. Hence, we can calculate the synchronizability
$R$ under different $\delta$. In Fig.5, synchronizability of the
spatial network is enhanced in some specifical exponent $\delta$.
Likewise, there is also an optimal $\delta$ for synchronizability
which is $1.5$ approximately when the network is large enough.

\begin{figure}
        \includegraphics[width=4.2cm]{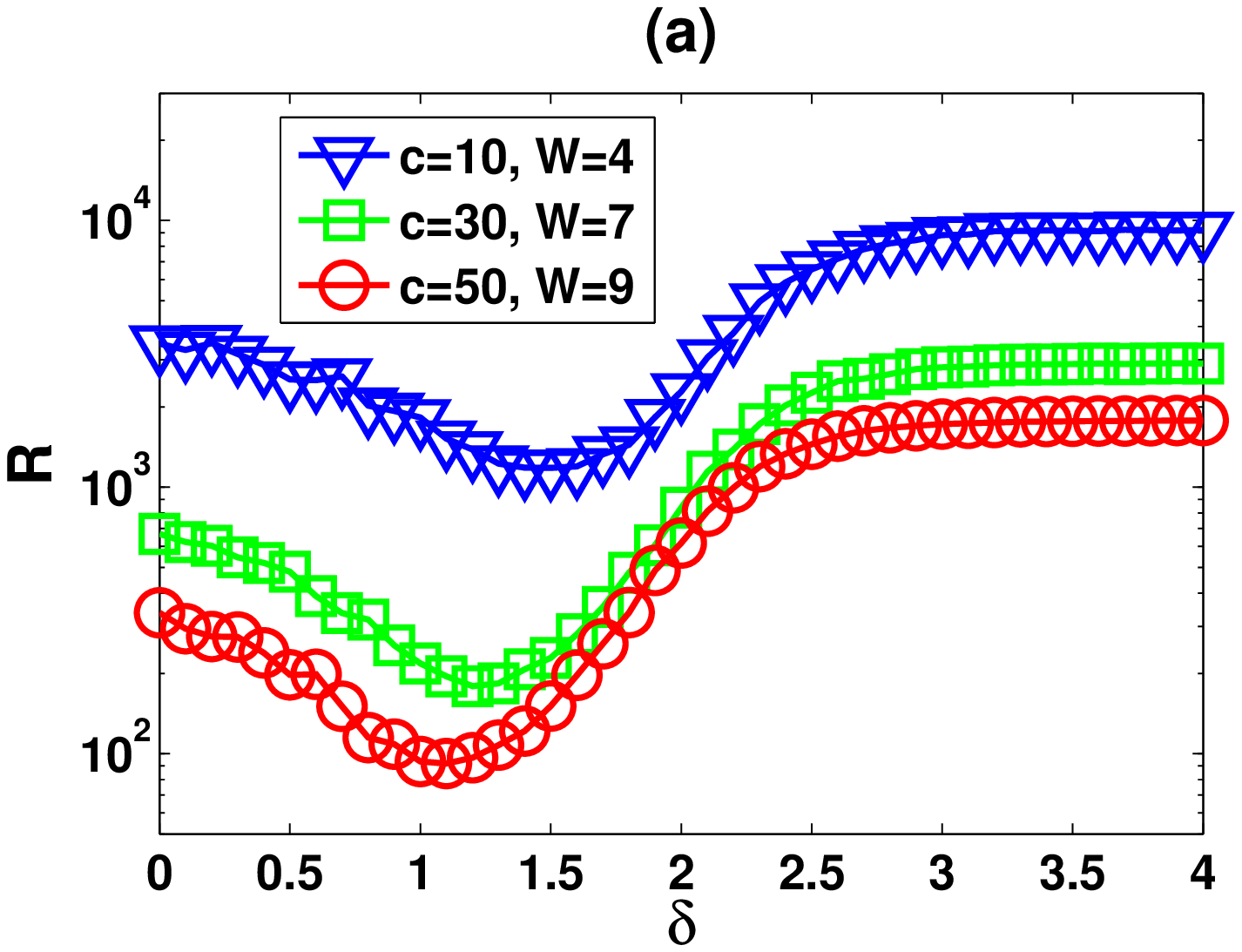}
        \includegraphics[width=4.2cm]{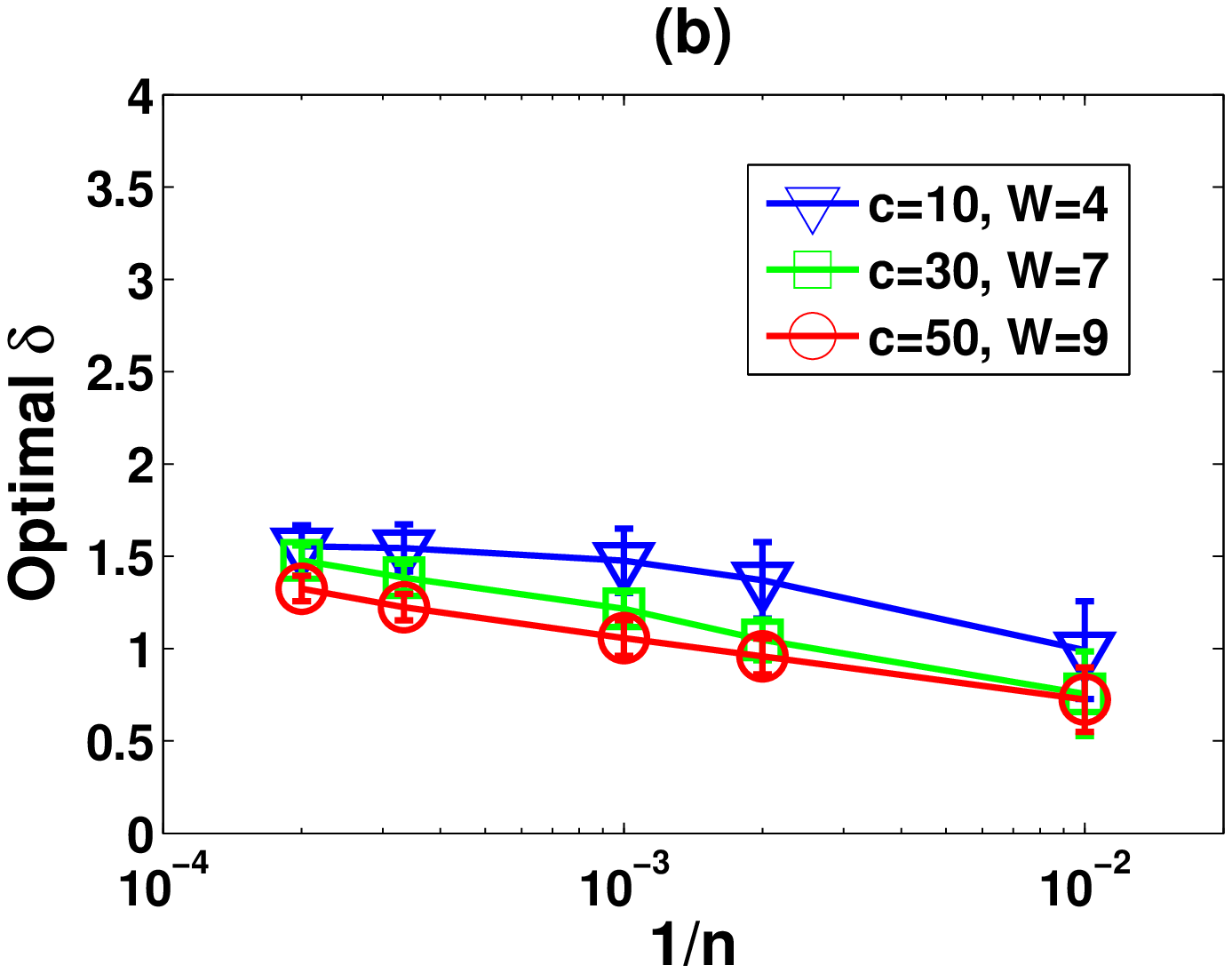}
\caption{(Color online) The effect of distance coarse graining on
the synchronizability of the binary spatial networks. (a) is the
synchronizability $R$ for different $\delta$ under different coarse
graining scale, where $N=1000$. (b) shows how the optimal $\delta$
for synchronizability changes when the network gets larger. The
results are under 50 independent realization.}
\end{figure}

\subsection{The Effect on Percolation}
We have known that there are two different ways to obtain the binary
spatial networks. The first is to project the weighted spatial
networks to an binary one, the second is to generate the binary
spatial networks directly by distance coarse graining procedure. To
begin with, we compared the percolation performance of the these two
kinds of networks. In percolation, we consider what happens with a
network if a random fraction $1-p$ of its edges is removed. In this
bond percolation problem, the giant connected component plays the
role of the percolation cluster, which may be destroyed by
decreasing $p$\cite{RMP1275}. Obviously, different $\delta$ in
spatial networks will result in different critical parameter
$p_{c}$. The smaller the critical parameter $p_{c}$ is, the better
the networks perform in percolation. The results for the percolation
performance of the these two kinds of binary spatial networks are
shown in Fig.6.

\begin{figure}
  \center
   \includegraphics[width=4.2cm]{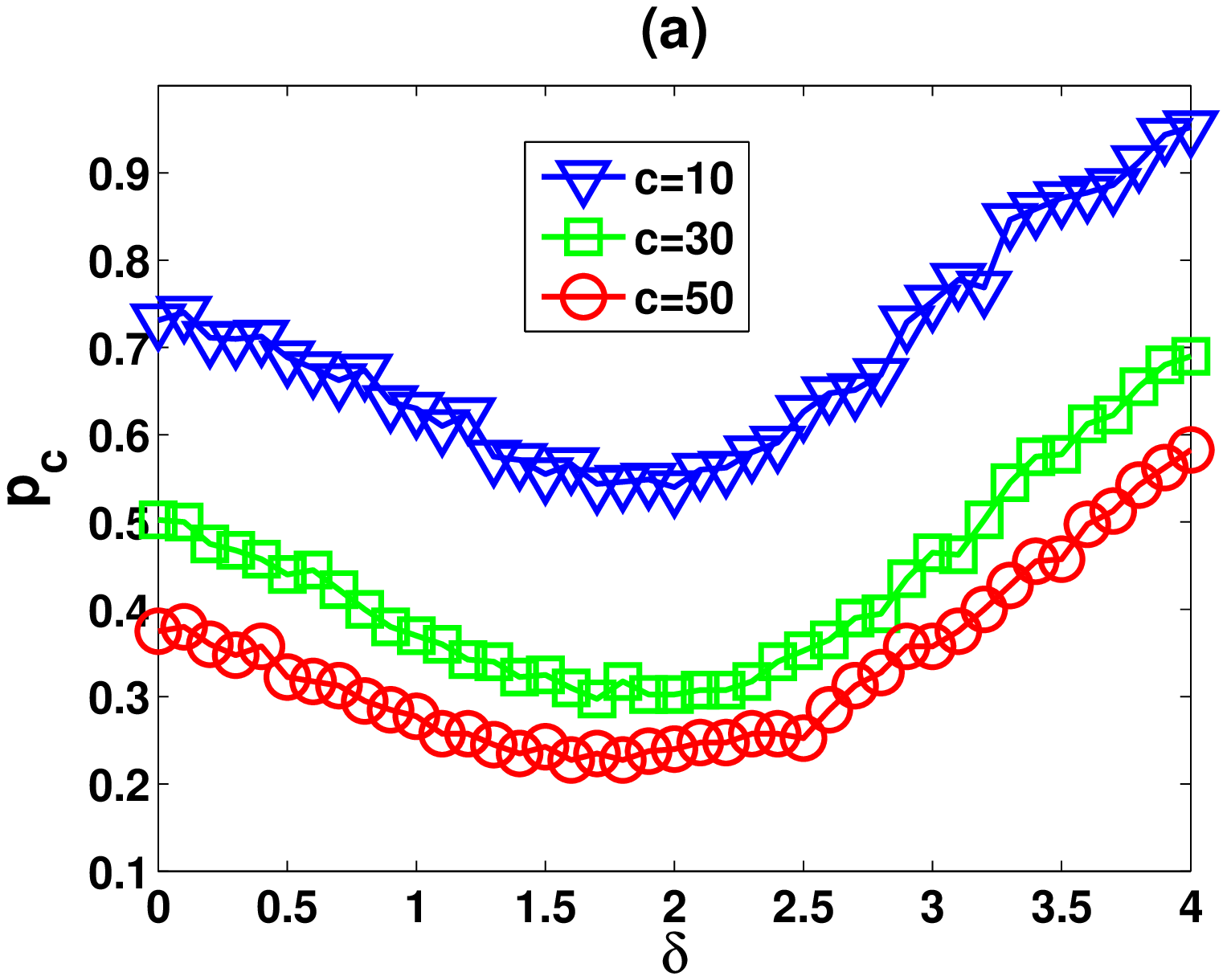}
   \includegraphics[width=4.2cm]{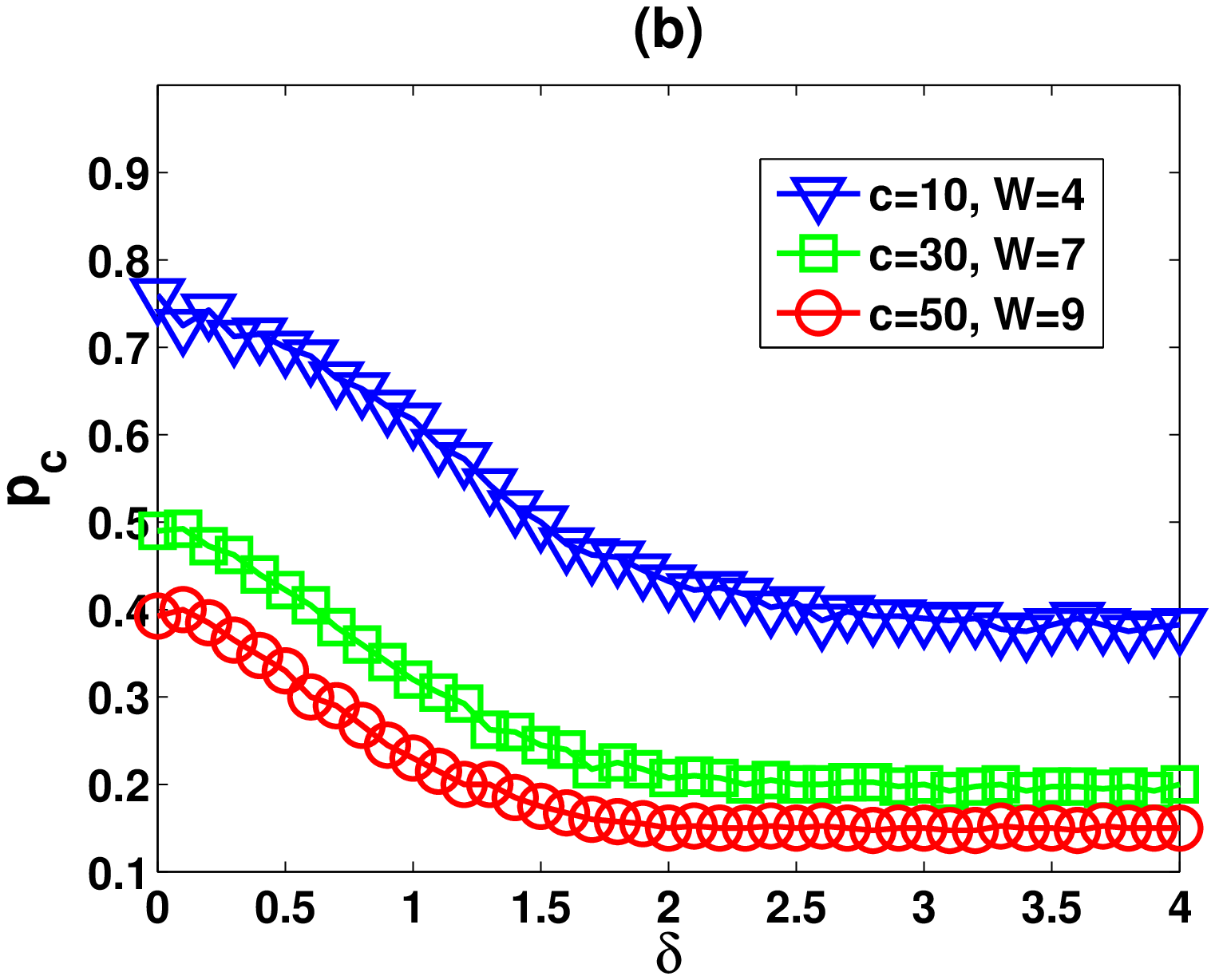}
  \caption{(Color online) The results for the percolation performance
of (a)projected binary spatial networks and (b) coarse grained
binary spatial networks. Here, the network size is $N=500$. The
results are under 50 independent realization. }
\end{figure}

From Fig.6, we can see that the projected binary spatial networks
have an optimal $\delta$ for percolation while coarse grained binary
spatial networks do not. Actually, when projecting the weighted
spatial networks, lots of energy gets lost. So in the projected
binary spatial, the number of total links become smaller as $\delta$
gets larger\cite{EPL58002}. This is why the projected binary spatial
networks have an optimal $\delta$ for percolation. Actually, when
designing the binary spatial network, it is supposed to constrained
by the total cost. In the percolation, we show that adding the total
energy constraint by the distance coarse graining, the spatial
network may perform entirely different in some functions. It
indicates the distance coarse graining is necessary when analyzing
some dynamics on spatial networks.

\section{Conclusion}

The complex networks has been a hot topic in science for more than
ten years. Works in this field are based on the topology of the
networks. So far, many empirical works claim that the distance
distribution of real networks obeys power-law distribution. In
theoretical modeling aspect, researches begin to pay attention to
these spatial networks. Very recently, these spatial networks are
constructed with power-law distance distribution and under total
energy restriction. This kind of spatial networks can reflect the
trade off property of real networks between the efficiency (distance
power-law distribution) and the total cost.

Studying the dynamic on spatial networks is of great significance,
which can lead to the enhancement of networks' specifical function
by choosing the proper exponent $\delta$. In previous works, authors
found that there exist stable optimal power-law indexes $\delta$ for
minimum shortest path, traffic process and navigation. With these
understandings, people may obtain the principle to design the
effective transport networks. In this paper, we study the
percolation and synchronizability in such binary spatial networks
with coarse grained distance. we find that synchronizability can
also be optimized by a typical $\delta$ while no optimal $\delta$
exists for percolation in such spatial network model.

In most case of real lives and empirical researches, the distances
are estimate approximately. In other words, people incline to regard
a range of distance as a typical distance. How this coarse graining
procedure affects the optimal $\delta$ is also studied in this
paper. Our results show that the distance coarse graining procedure
will make the optimal exponent $\delta$ in power-law distance
distribution shift to smaller values for all of average shortest
path, traffic process and navigation. Interestingly, when the
network is large enough, the effect of distance coarse graining can
be ignored. As the real networks, say the transport networks, is
usual of relatively large size, the result indicates that the
optimal index $\delta$ still works in designing principles for the
optimal real transport networks. These results above indicate that
the distance coarse graining can be used as a universal way to
generate the binary spatial model with its total cost constraint
satisfied and its power-law distance distribution preserved
effectively. Moreover, investigation of functions of spatial related
networks with coarse grained distance can be an interesting
extension.

\section*{Acknowledgement}
The authors would like to thank professor Shlomo Havlin for many
helpful suggestions. This work is partially supported by the 985
Project and NSFC under the grants No. 70771011 and No. 60974084.


\begin{thebibliography}{99}
\bibitem{explain} Actually, probability $Pr(u,v)$ of two nodes (u and v) to have a long range
connection is different from the probability density function (PDF)
$P(r)$ of the distance distribution. Specifcally, in 1-dimensional
space, $P(r)\propto Pr(u,v)$. in 2-dimensional space, the number of
nodes which have distance $r$ from a given node is proportional to
$r$, so the PDF is $P(r)\propto r\cdot Pr(u,v)$.
\end{thebibliography}
\end{document}